\documentclass{PoS}

\title{Inclusive and diffractive dijet photoproduction in UPCs at the LHC in NLO QCD}

\ShortTitle{Inclusive and diffractive dijet photoproduction in UPCs at the LHC in NLO QCD}

\author{Vadim Guzey\thanks{Work supported by RFBR through the grant 17-52-12070.}\\
  National Research Center ``Kurchatov Institute'',
  Petersburg Nuclear Physics Institute (PNPI),
  Gatchina, 188300, Russia\\
  Department  of  Physics,  University  of  Jyv\"askyl\"a, P.O.\
  Box 35, 40014  University  of  Jyv\"askyl\"a,  Finland\\
  Helsinki Institute of Physics, P.O.\  Box  64,  00014
  University  of  Helsinki,  Finland\\
  E-mail: \email{guzey\_va@nrcki.pnpi.ru}}

\author{\speaker{Michael Klasen}\thanks{Work supported by the DFG through the
  grant KL 1266/9-1.}\\
  Institut f\"ur Theoretische Physik,
  Westf\"alische Wilhelms-Universit\"at M\"unster,
  Wilhelm-Klemm-Stra\ss{}e 9,
  48149 M\"unster, Germany\\
  E-mail: \email{michael.klasen@uni-muenster.de}}

\abstract{We present a next-to-leading order QCD calculation of inclusive
  dijet photoproduction in ultraperipheral Pb-Pb collisions at the LHC and show
  that the results agree very well with various kinematic distributions measured
  by the ATLAS collaboration. The effect of including these data in nCTEQ or EPPS16
  nuclear parton density functions (nPDFs) is then studied using the Bayesian
  reweighting technique. For an assumed total error of 5\% on the final data, its
  inclusion would lead to a significant reduction of the nPDF uncertainties of up to
  a factor of two at small values of the parton momentum fraction. As an outlook,
  we discuss future analyes of diffractive nPDFs, which are so far completely unknown.}

\FullConference{
European Physical Society Conference on High Energy Physics - EPS-HEP2019 -\\
			10-17 July, 2019\\
			Ghent, Belgium}

\newcommand{\Pomeron}{I\!\!P}

\begin{document}

\section{Motivation}

Ultraperipheral collisions (UPCs) of relativistic ions are defined by a
large impact parameter $b$ that exceeds the sum of the nuclear radii
$R_A$. At these large distances, short-range strong nuclear forces
are suppressed. The nuclei interact instead electromagnetically through
long-range photon exchanges in $\gamma\gamma$ and $\gamma A$ reactions,
in particular when they are as heavily charged as lead ions ($Z=82$)
\cite{Baltz:2007kq}.
Interesting examples of physics processes in UPCs include quarkonium and
dilepton pair production, light-by-light scattering and searches for
physics beyond the Standard Model. Here we focus on inclusive dijet
photoproduction in Pb-Pb collisions at the LHC, which has recently been
observed and analysed by the ATLAS collaboration \cite{ATLAS:2017kwa}.

Apart from the novelty of the observation, this process is particularly
interesting for future constraints on nuclear parton distribution
functions (PDFs) $f_{j/A}(x,Q^2)$ \cite{Kovarik:2015cma,Eskola:2016oht}.
They can be modeled from their bare proton counterparts $f_{j/p}(x,Q^2)$
with a multiplicative factor $R_j^A(x,Q^2)$, which captures the nuclear
modifications. Depending on the region in the momentum fraction $x$,
different effects have been observed. At low $x$, the shadowing
suppression can be interpreted as the absorption by surface nucleons
of the virtual photon probing the nucleus after fluctuating into
$q\bar{q}$ dipoles. Shadowing is compensated at intermediate $x$ by
antishadowing as imposed by the momentum sum rule. At large $x$,
nuclear PDFs are again reduced by the EMC effect, interpreted in
various ways like valence quark suppression due to nuclear binding,
pion exchange, quark clusters, short-range correlations, etc. At
very large $x$, Fermi motion of the nucleons leads to nuclear
enhancement. The extraction of nuclear PDFs suffers from large
uncertainties, in particular for gluons at small $x$, so that the
inclusion of LHC and future EIC data is very important
\cite{Aschenauer:2017oxs,Klasen:2017kwb} .

\section{Inclusive dijet photoproduction at the LHC}

\begin{figure}
  \centering
  \includegraphics[width=0.8\textwidth]{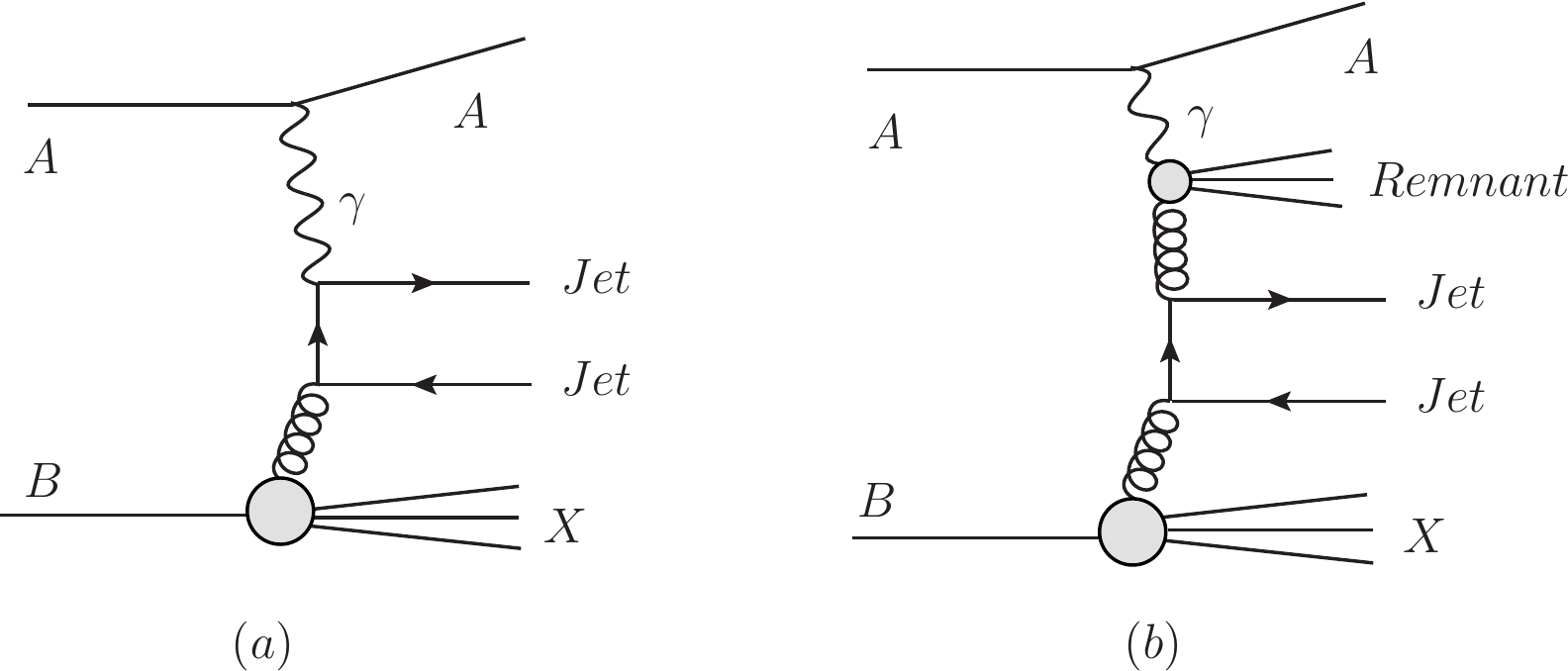}
  \caption{Direct (left) and resolved (right) photoproduction of
    dijets in ultraperipheral collisions of nuclei $A$ and $B$
    at the LHC. \label{fig:1}}
\end{figure}

A potentially interesting novel process in this respect is inclusive
dijet photoproduction in UPCs at the LHC, which has recently been
observed and analysed by the ATLAS collaboration \cite{ATLAS:2017kwa}.
We have computed this process in NLO QCD \cite{Guzey:2018dlm}, based
on previous work on inclusive jet \cite{Klasen:1994bj} and dijet
\cite{Klasen:1995xe}, real and virtual \cite{Klasen:1997jm}
photoproduction in $ep$ collisions at HERA (for a review see
\cite{Klasen:2002xb}). As shown in Fig.\ \ref{fig:1}, direct (left)
and resolved (right) photons contribute to these processes.

The differential hadronic cross sections
\begin{equation}
d\sigma(AB\to AB+{\rm 2\,jets}+X)=
\sum_{a,b} \int \!dy\! \int \!dx_{\gamma} \!\int\! dx_A\, f_{\gamma/A}(y)f_{a/\gamma}(x_{\gamma},\mu_f^2)f_{b/B}(x_A,\mu_f^2) d\hat{\sigma}(ab \to {\rm jets})\nonumber
\end{equation}
are related to those of partons $a$ and $b$, $d\hat{\sigma}(ab \to {\rm jets})$,
by the photon flux $f_{\gamma/A}(y)$ and PDFs $f_{a/\gamma}(x_{\gamma},\mu_f^2)$,
where the former are well described by
\begin{equation}
 f_{\gamma/A}(y)=\frac{2 \alpha Z^2}{\pi}\frac{1}{y} \left[\zeta K_0(\zeta)K_1(\zeta)-\frac{\zeta^2}{2}(K_1^2(\zeta)-K_0^2(\zeta)) \right] \nonumber
\end{equation}
for a relativistic pointlike charge $Z$ with $\zeta\!=ym_pb_{\min}$,
assuming no strong interactions for $b>b_{\min}=2.1R_{\rm Pb}=14.2$\,fm.
The latter are taken from the GRV NLO parameterisation \cite{Gluck:1991jc},
while we adopt nCTEQ15 nuclear PDFs \cite{Kovarik:2015cma} and estimate
their uncertainty by summing over independent eigenvectors, $\Delta
\sigma =\frac{1}{2} \sqrt{ \sum_{k=1}^{31}\left(\sigma(f_k)\!-\!\sigma(f_{k+1})
\right)^2}$. The renormalisation and factorisation scales are set to
$\mu_r=\mu_f=2E_{T,1}$, where the points of fastest convergence of the
perturbative seires and of minimal scale sensitivity coincide. Jets are
defined with the anti-$k_T$ algorithm and distance parameter $R=0.4$,
transverse energies $E_{T,1}>20$ GeV, $E_{T,2}>15$ GeV, $H_T=\sum_i E_{T,i}>35$
GeV, rapidities $|\eta_{1,2}|<4.4$ and a combined jet mass $m_{\rm jets}>35$ GeV.

\begin{figure}[b]
  \centering
  \includegraphics[width=0.49\textwidth]{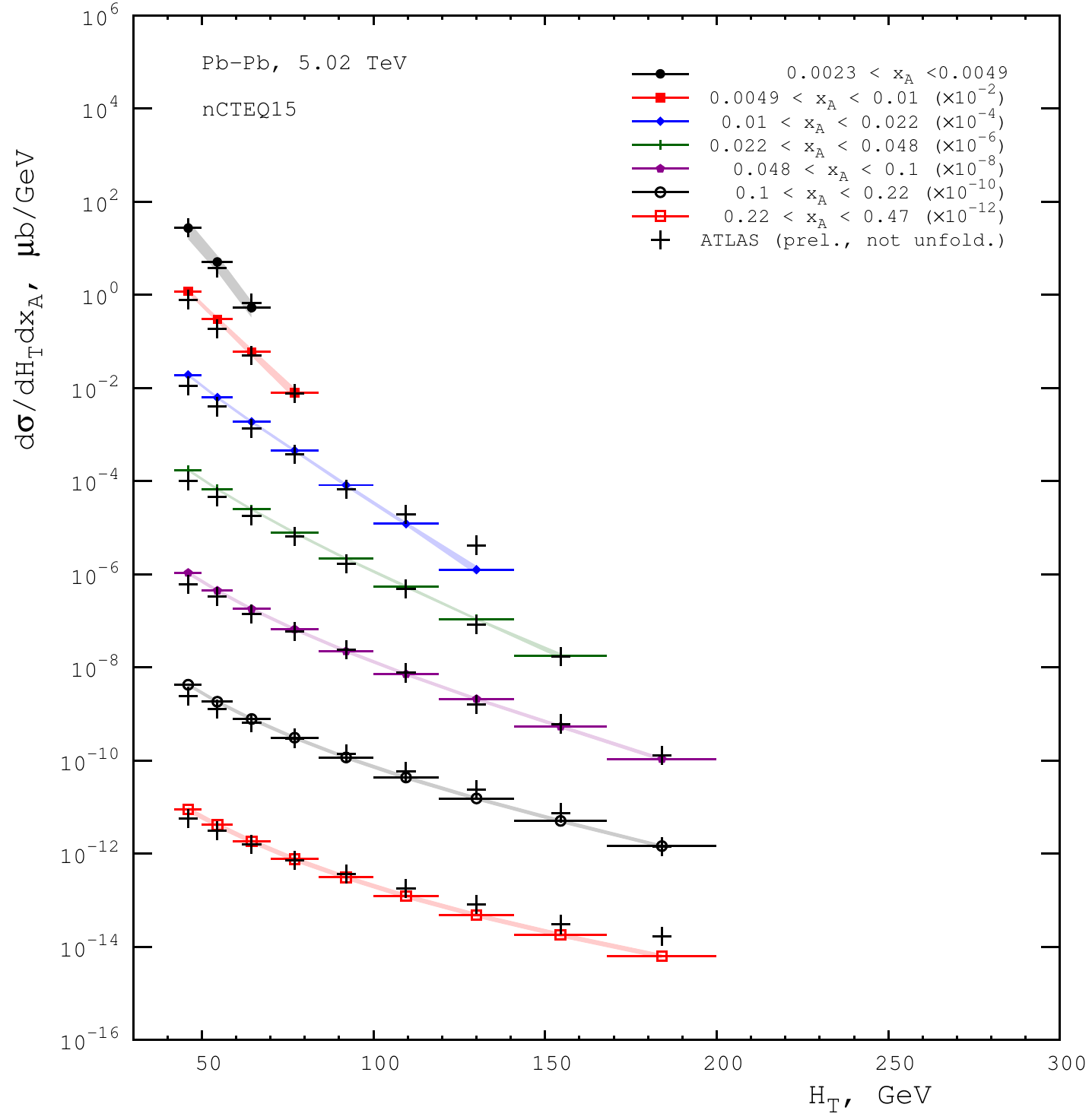}
  \includegraphics[width=0.49\textwidth]{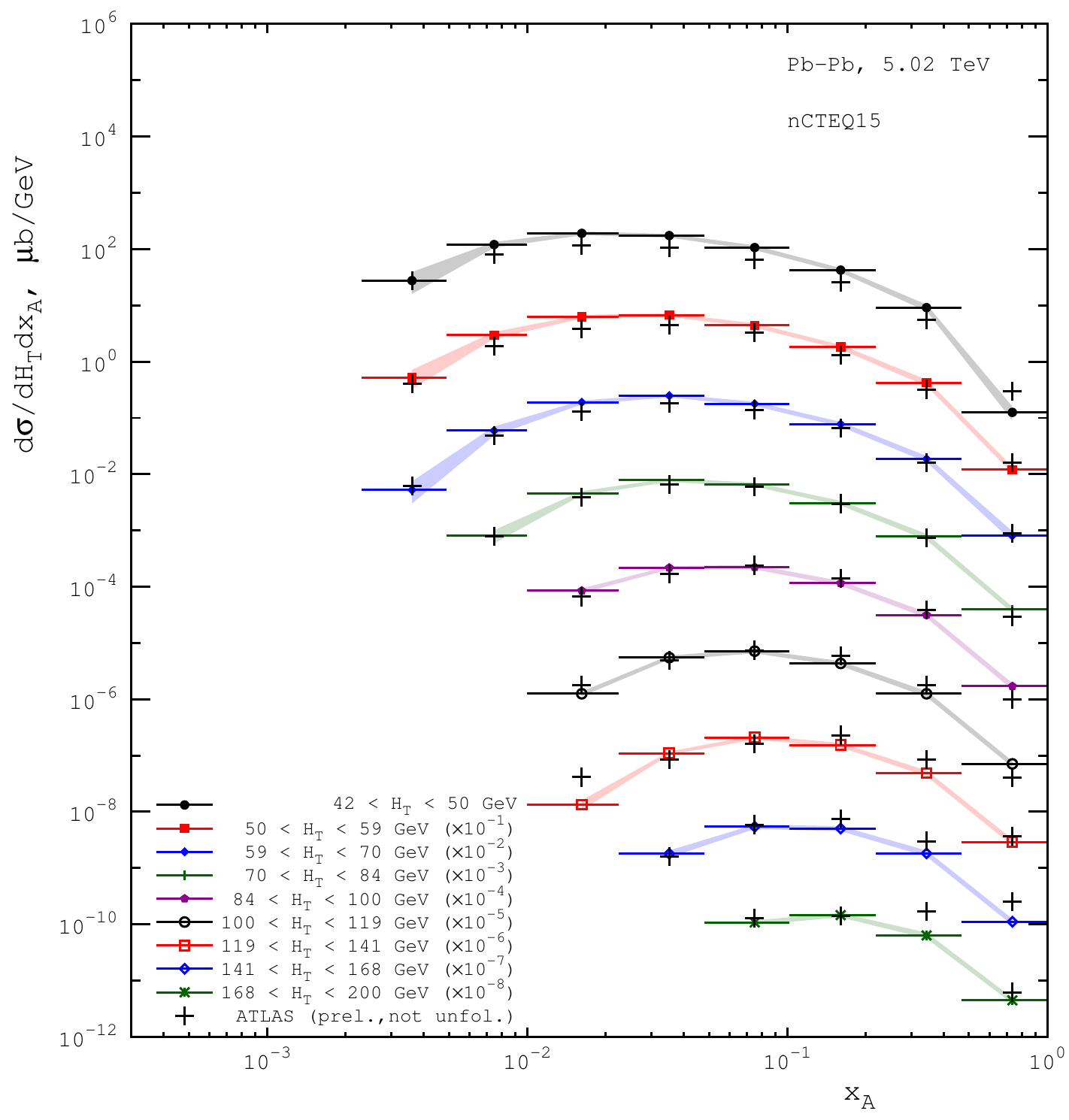}
  \caption{Inclusive dijet photoproduction at the LHC as measured by ATLAS,
    compared to our NLO QCD calculations. Shown are the double-differential
    total transverse energy (left) and parton momentum fraction distributions
    in the target lead ion (right) \cite{Guzey:2018dlm}. \label{fig:2}}
\end{figure}

The comparison of our calculations with the - unfortunately still preliminary
- ATLAS data \cite{ATLAS:2017kwa} is shown in Fig.\ \ref{fig:2}.
On a logartihmic scale, we find excellent agreement not only in the total
transverse energy ($H_T$, left) and parton momentum fraction distributions
in the lead ion ($x_A$, right), but also for the parton momentum fraction
distributions in the photon ($z_\gamma$, not shown). Note, however, that the
data have not yet been unfolded for detector response.\\

\begin{figure}[t]
  \centering
  \includegraphics[width=\textwidth]{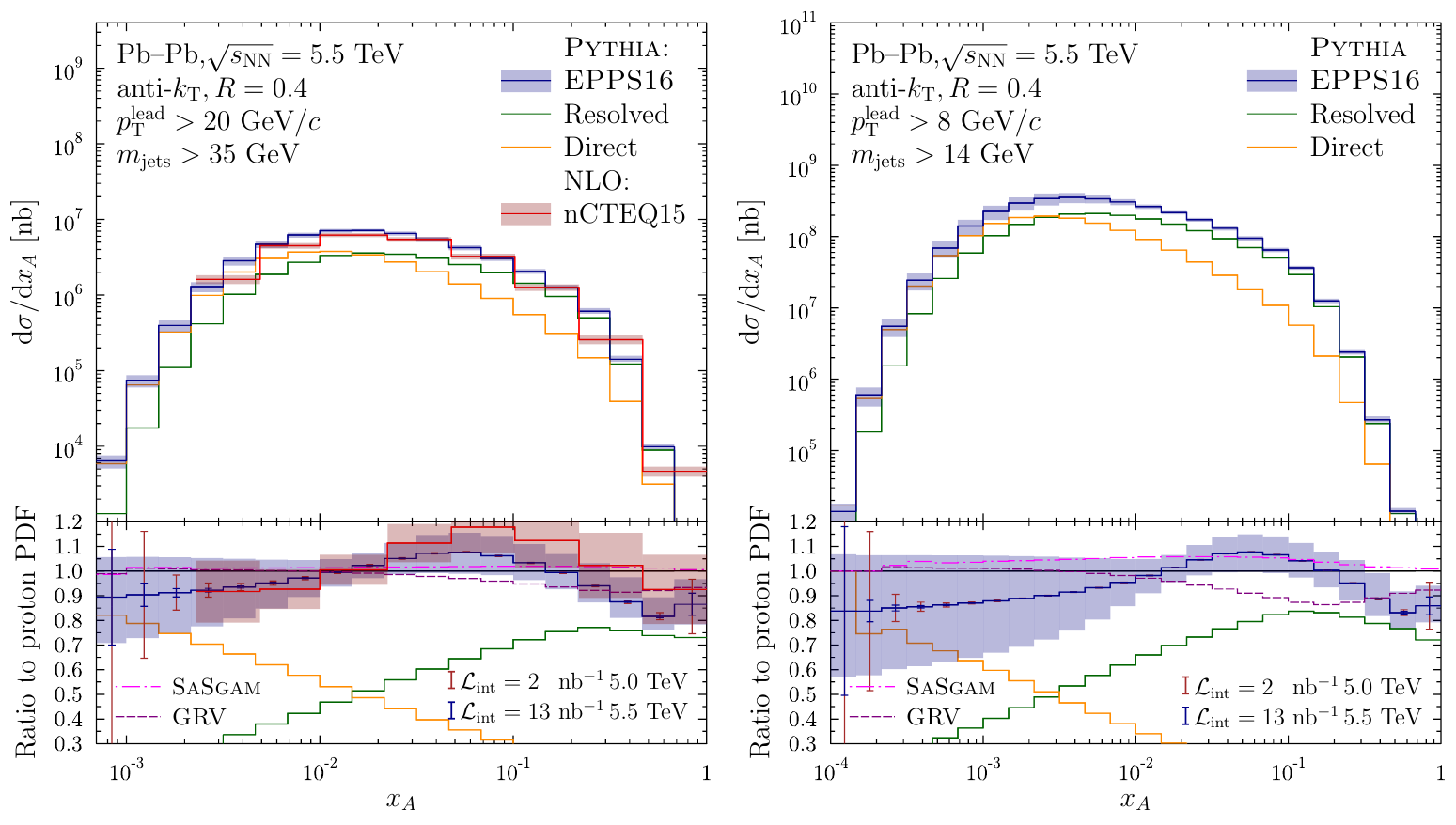}
  \caption{Single-differential parton momentum fraction distributions integrated
    over $H_T$ for the HL-/HE-LHC and 5.5 TeV centre-of-mass energy with original
    (left) and low-$x_A$-extended ATLAS acceptance (right) \cite{Citron:2018lsq}.
    Also shown are simulations with PYTHIA 8, direct and resolved contributions
    separately, and the dependence on the photon PDFs \cite{Helenius:2018bai}.
    \label{fig:3}}
\end{figure}

For the high-luminosity (HL) and high-energy (HE) LHC community study, we
have updated our predictions from 5.02 to 5.5 TeV centre-of-mass energy
per nucleon \cite{Citron:2018lsq}. The results are shown in Fig.\ \ref{fig:3}.
We observe a large potential for improvement in the nuclear shadowing region,
in particular if the ATLAS modifies the acceptance from the current transverse
energy cuts (left) to lower values (right). As one can see, the resolved photon
PDF sensitivity resides mostly at large $x_A$ corresponding to small $z_\gamma$
or low $E_T$, while the direct process dominates at small $x_A$.

\section{Bayesian reweighting}

\begin{figure}
  \centering
  \includegraphics[width=\textwidth]{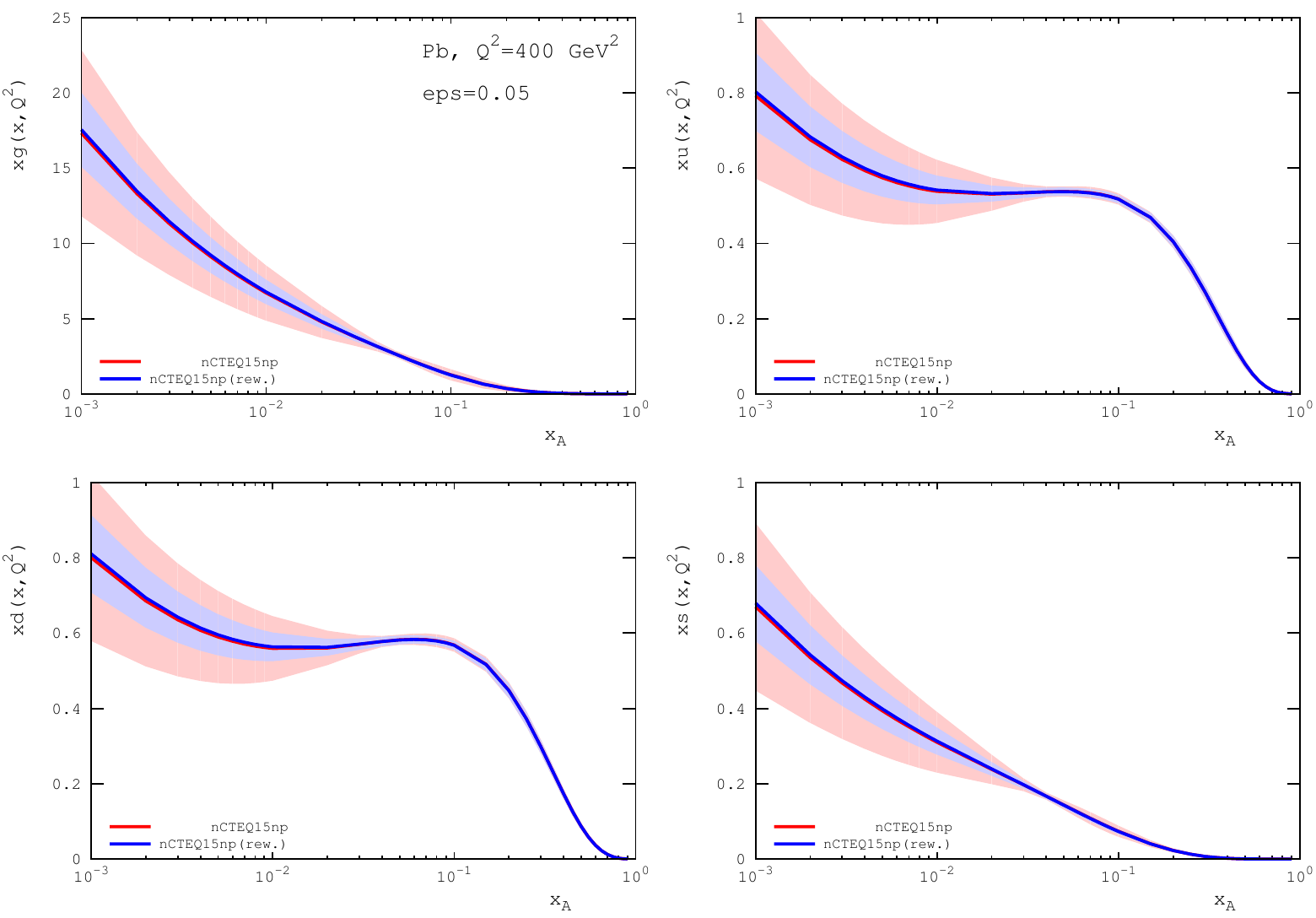}
  \caption{The gluon, $u$-quark, $d$-quark, and $s$-quark nCTEQ15np nPDFs as a function
    of $x_A$ at $Q^2=400$ GeV$^2$ with (blue, inner band) and without (red, outer band)
    the Bayesian reweighting for an assumed total experimental uncertainty of
    $\epsilon=0.05$ \cite{Guzey:2019kik}. \label{fig:4}}
\end{figure}

Using our NLO QCD calculations presented in the previous section, we
went on to study the impact of dijet photoproduction data at the LHC on
future determinations of nuclear PDFs \cite{Guzey:2019kik}. Denoting
the central fits of the nCTEQ15 \cite{Kovarik:2015cma} and EPPS16
\cite{Eskola:2016oht} analyses by $f^0_{j/A}$ for parton $j$ and nucleus
$A$ and the error sets by $f^{i\pm}_{j/A}$ ($i=1\,...\,2N$ with $N=16$ for
nCTEQ15, based on CTEQ6.1M proton PDFs, and $N=20+28$ for nuclear + proton
PDF uncertainties in EPPS16), we
produced replicas $k=1\,...\,N_{\rm rep}$ with $N_{\rm rep}=10,000$ through
\begin{equation}
f_{j/A}^k(x,Q^2)=f_{j/A}^0(x,Q^2)+\frac{1}{2}\sum_{i=1}^{N} \left[f_{j/A}^{i+}(x,Q^2)-f_{j/A}^{i-}(x,Q^2)\right] R_{ki}
\end{equation}
with a normally distributed random number $R_{ki}$ ($\mu=0,\sigma=1$)
as well as pseudodata from our NLO QCD prediction for $d\sigma^0/dx_A$
with the central PDFs $f^0_{j/A}$ for $N_{\rm data}=9$ bins in $x_A$. We
then evaluated the test function
\begin{equation}
\chi_k^2=\sum_{j=1}^{N_{\rm data}} \frac{(d\sigma^0/dx_A-d\sigma^k/dx_A)^2}{\sigma_j^2} \nonumber
\end{equation}
with assumed uncertainties $\sigma_j=\epsilon d\sigma^0/dx_A$ for different
assumptions on the data precision $\epsilon=0.05\,...\, 0.2$. This allowed
us to obtain reweighted nPDFs
\begin{eqnarray}
\langle f_{j/A}(x,Q^2) \rangle_{\rm new} &=& \frac{1}{N_{\rm rep}} \sum_{k=1}^{N_{\rm rep}} w_k f_{j/A}^k(x,Q^2)
\end{eqnarray}
and their uncertainties
\begin{eqnarray}
\delta \langle f_{j/A}(x,Q^2) \rangle_{\rm new} &=& \sqrt{\frac{1}{N_{\rm rep}} \sum_{k=1}^{N_{\rm rep}} 
w_k \left(f_{j/A}^k-\langle f_{j/A}(x,Q^2) \rangle_{\rm new}\right)^2}
\end{eqnarray}
from the weights
\begin{equation}
w_k=\frac{e^{-\frac{1}{2}\chi_k^2/T}}{\frac{1}{N_{\rm rep}} \sum_i^{N_{\rm rep}}e^{-\frac{1}{2}\chi_i^2/T}},
\end{equation}
where $\sum_k w_k=N_{\rm rep}$ and the tolerances were $T=35$ and 52 for nCTEQ15
and EPPS16, respectively. The effective numbers of contributing replicas are then
\begin{equation}
N_{\rm eff}=\exp \left[\frac{1}{N_{\rm rep}} \sum_k^{N_{\rm rep}} w_k \ln(N_{\rm rep}/w_k) \right].
\end{equation}
They are listed in Tab.\ \ref{tab:1}.
The impact of the final ATLAS data with an assumed total uncertainty of
$\epsilon=0.05$ on the nPDFs can be deduced from Fig.\ \ref{fig:4}.
In this case, the uncertainty of the nCTEQ15np fit, which does not use pion
data from RHIC, is reduced by about a factor of two, in particular at low $x_A$.

\begin{table}[t]
  \caption{Effective number of contributing replicas in our nPDF reweighting study. \label{tab:1}}
\begin{center}
\begin{tabular}{|c|c|c|c|}
\hline
$\epsilon$ & $N_{\rm eff}$(nCTEQ15) &  $N_{\rm eff}$(nCTEQ15np) &  $N_{\rm eff}$(EPPS16) \\
\hline
0.05 & 4407 & 3982  & 5982 \\
0.1 &  7483 & 7742  & 8727 \\
0.15 & 8870 & 9107  & 9555 \\
0.2  & 9464 & 9607  & 9818 \\
\hline
\end{tabular}
\end{center}
\end{table}

\section{Diffractive dijet photoproduction}

A completely novel set of PDFs, namely diffractive nuclear PDFs
$f^{D(4)}_{b/A}(x_{\Pomeron},z_{\Pomeron},t,\mu^2)$, appears in cross
sections of diffractive dijet photoproduction
\begin{equation}
d \sigma 
=\sum_{a,b} \int 
 dt \int 
 dx_{\Pomeron}
\int 
 dz_{\Pomeron} \int
 dy \int 
 dx_{\gamma}
 f_{\gamma/A}(y)  f_{a/\gamma}(x_{\gamma},\mu_f^2) f^{D(4)}_{b/A}(x_{\Pomeron},z_{\Pomeron},t,\mu_f^2)
d \hat{\sigma}_{ab \to {\rm jets}}^{(n)}
\end{equation}
with intact (at most excited) nuclei and/or large rapidity gaps
on both sides of the event. Diffractive nuclear PDFs can be theoretically
defined as conditional leading-twist distributions of partons $b$
in nuclei $A$ in terms of the light-cone momentum fraction
$z_{\Pomeron}$ at the resolution scale $\mu_f$, provided  that the nucleus
undergoes diffractive scattering characterised by the light-cone momentum
fraction loss $x_{\Pomeron}$ and the invariant momentum transfer squared $t$. 
The leading-twist model of nuclear shadowing \cite{Frankfurt:2011cs}, which
is based on a generalisation of Gribov-Glauber theory, QCD factorisation
theorems and information on diffractive processes at HERA \cite{Klasen:2004qr,%
Guzey:2016awf}, predicts a significant suppression of nuclear diffractive PDFs
\begin{equation}
f^{D(4)}_{b/A}(x_{\Pomeron},z_{\Pomeron},t,\mu_f^2)=R_b(x_{\Pomeron},z_{\Pomeron},\mu_f^2) f^{D(4), {\rm IA}}_{b/A}(x_{\Pomeron},z_{\Pomeron},t,\mu_f^2) 
\label{eq:4.2}
\end{equation}
at low $x_A=x_{\Pomeron}z_{\Pomeron}$ compared to the impulse approximation (IA).
Note that Eq.\ (\ref{eq:4.2}) breaks in principle the phenomenological factorisation
of diffractive PDFs into the product of a Pomeron ($\Pomeron$) flux and Pomeron
PDFs. However, the shadowing suppression $R_b(x_{\Pomeron},z_{\Pomeron},\mu^2)$ depends
only weakly on the parton flavor $b$, the scale $\mu_f$, $z_{\Pomeron}$ and $x_{\Pomeron}$
and can in practice be approximated by a factor of 0.15.

In a recent study, we have made predictions in NLO QCD for diffractive
dijet photoproduction in pp, p-Pb and Pb-Pb collisions at the LHC
\cite{Guzey:2016tek}. Distributions for the latter at a center-of-mass
\begin{figure}
  \centering
  \includegraphics[width=\textwidth]{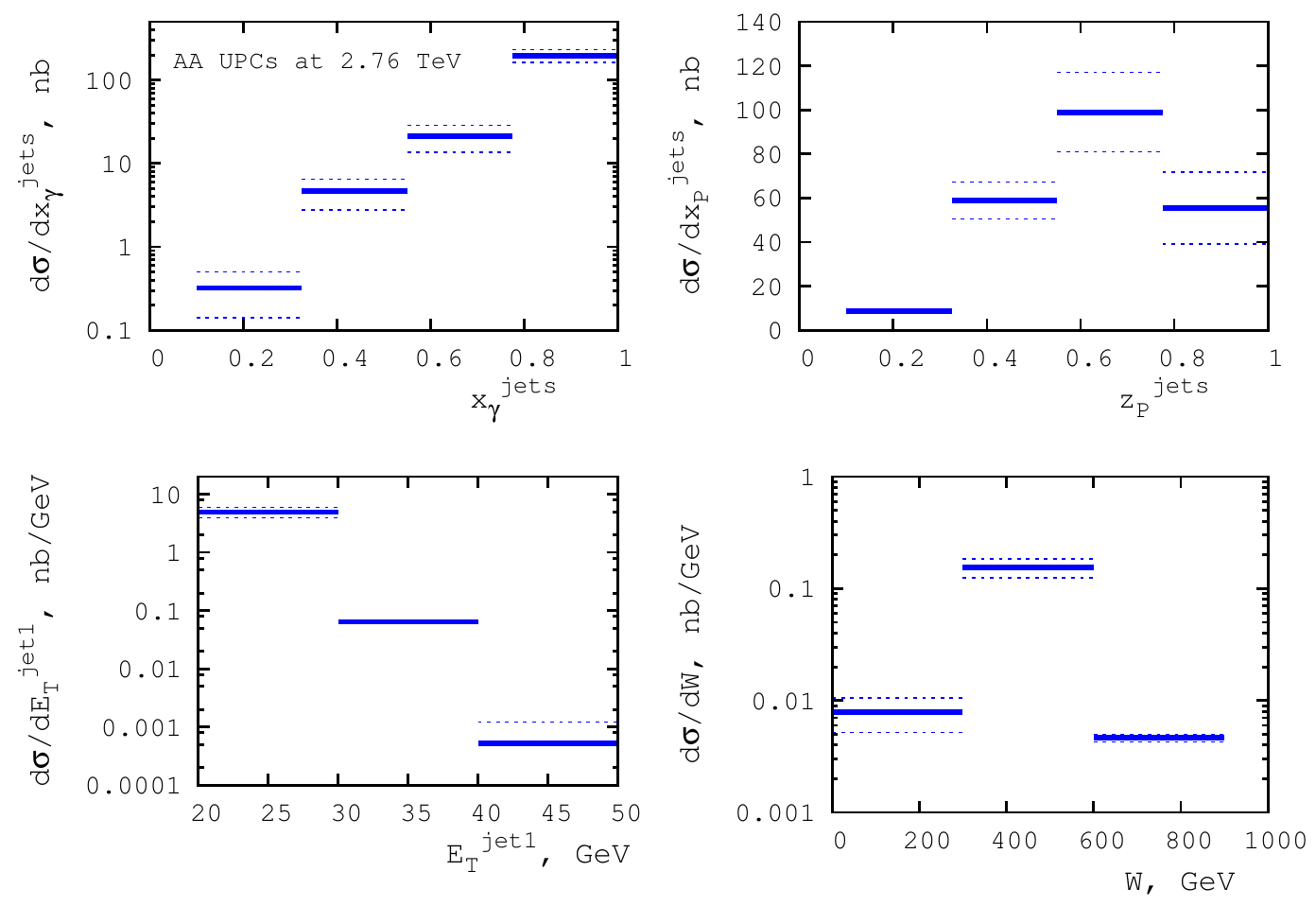}
  \caption{Differential cross sections for diffractive photoproduction of dijets
    $d \sigma(AA\to A+2{\rm jets}+ X^{\prime}+A)$ in Pb-Pb UPCs at
    $\sqrt{s_{NN}}=2.76$ TeV.\label{fig:5}}
\end{figure}
energy per nucleon of $\sqrt{s_{NN}}=2.76$ TeV are shown in Fig.\ \ref{fig:5}.
Approximate results for p-Pb and pp collisions and can be obtained by a
simple rescaling with $A$ based on the approximate relation
$f_{b/A}^{D(3)}(x_{\Pomeron},z_{\Pomeron},\mu_f^2) \approx A/2  f_{b/p}^{D(3)}(x_{\Pomeron},z_{\Pomeron},\mu_f^2)$
between the nuclear and proton diffractive PDFs (integrated over the momentum
transfer $t$) and the fact that the Pb-Pb cross section receives contributions
of both nuclei, while the p-Pb cross section is dominated by the
photon-from-nucleus contribution. 

It is well known from studies of diffractive photoproduction of dijets in $ep$
scattering at HERA that collinear factorisation for this process is broken,
{\it i.e.} NLO QCD calculations overestimate the measured cross sections by
almost a factor of two. The pattern of this factorization breaking remains
unknown and presents one of the outstanding questions in this field 
\cite{Klasen:2004qr,Guzey:2016awf}, but it would in principle of course also
apply to diffractive dijet photoproduction in UPCs at the LHC \cite{Guzey:2016tek}.

\section{Conclusion}

In conclusion, we have presented an NLO QCD analysis of dijet photoproduction
in UPCs at the LHC, where short-range strong interactions are suppressed.
On a logarithmic scale, our calculations agreed very well with
preliminary ATLAS data, which unfortunately must still be unfolded for
detector effects. In a Bayesian reweighting study, we showed that the
final data have the potential to reduce the uncertainties of nuclear PDFs,
in particular in the shadowing region at small parton momentum fractions,
by about a factor of two. If the nuclei on both sides of the interaction
stay intact or have large rapidity gaps to the central hard event, an
interesting novel quantity, namely diffractive nuclear PDFs can be extracted,
in particular, but not only, from diffractive dijet photoproduction, and QCD
factorisation breaking in these processes can be analysed in detail.


\begin{thebibliography}{99}

\bibitem{Baltz:2007kq}
  A.~J.~Baltz {\it et al.},
  Phys.\ Rept.\  {\bf 458} (2008) 1.

\bibitem{ATLAS:2017kwa}
  The ATLAS collaboration [ATLAS Collaboration],
  ATLAS-CONF-2017-011.

\bibitem{Kovarik:2015cma}
  K.~Kovarik {\it et al.},
  Phys.\ Rev.\ D {\bf 93} (2016)  085037;
%
  K.~Kovarik, P.~M.~Nadolsky and D.~E.~Soper,
  arXiv:1905.06957 [hep-ph];
%
  D.~B.~Clark {\it et al.} [nCTEQ Collaboration],
  arXiv:1909.00452 [hep-ph].

\bibitem{Eskola:2016oht}
  K.~J.~Eskola, P.~Paakkinen, H.~Paukkunen and C.~A.~Salgado,
  Eur.\ Phys.\ J.\ C {\bf 77} (2017) 163.

\bibitem{Aschenauer:2017oxs}
  E.~C.~Aschenauer, S.~Fazio, M.~A.~C.~Lamont, H.~Paukkunen and P.~Zurita,
  Phys.\ Rev.\ D {\bf 96} (2017) 114005.

\bibitem{Klasen:2017kwb}
  M.~Klasen, K.~Kovarik and J.~Potthoff,
  Phys.\ Rev.\ D {\bf 95} (2017) 094013;
%
  M.~Klasen and K.~Kovarik,
  Phys.\ Rev.\ D {\bf 97} (2018) 114013.

\bibitem{Guzey:2018dlm}
  V.~Guzey and M.~Klasen,
  Phys.\ Rev.\ C {\bf 99} (2019)  065202.

\bibitem{Klasen:1994bj}
  M.~Klasen, G.~Kramer and S.~G.~Salesch,
  Z.\ Phys.\ C {\bf 68} (1995) 113;
%
  M.~Klasen, G.~Kramer and M.~Michael,
  Phys.\ Rev.\ D {\bf 89} (2014)  074032.
  
\bibitem{Klasen:1995xe}
  M.~Klasen and G.~Kramer,
  Phys.\ Lett.\ B {\bf 366} (1996) 385;
%
  Z.\ Phys.\ C {\bf 72} (1996) 107;
%
  Z.\ Phys.\ C {\bf 76} (1997) 67;
%
  Eur.\ Phys.\ J.\ C {\bf 71} (2011) 1774;
%
  M.~Klasen, T.~Kleinwort and G.~Kramer,
  Eur.\ Phys.\ J.\ direct {\bf 1} (1998) 1.

\bibitem{Klasen:1997jm}
  M.~Klasen, G.~Kramer and B.~P\"otter,
  Eur.\ Phys.\ J.\ C {\bf 1} (1998) 261;
%
  T.~Biek\"otter, M.~Klasen and G.~Kramer,
  Phys.\ Rev.\ D {\bf 92} (2015) 074037.

\bibitem{Klasen:2002xb}
  M.~Klasen,
  Rev.\ Mod.\ Phys.\  {\bf 74} (2002) 1221.
  
\bibitem{Gluck:1991jc}
  M.~Gl\"uck, E.~Reya and A.~Vogt,
  Phys.\ Rev.\ D {\bf 46} (1992) 1973.

\bibitem{Citron:2018lsq}
  Z.~Citron {\it et al.},
  arXiv:1812.06772 [hep-ph].

\bibitem{Helenius:2018bai}
  I.~Helenius,
  PoS DIS {\bf 2018} (2018) 113.
  
\bibitem{Guzey:2019kik}
  V.~Guzey and M.~Klasen,
  Eur.\ Phys.\ J.\ C {\bf 79} (2019) 396.

\bibitem{Frankfurt:2011cs}
  L.~Frankfurt, V.~Guzey and M.~Strikman,
  Phys.\ Rept.\  {\bf 512} (2012) 255.

\bibitem{Klasen:2004qr}
  M.~Klasen and G.~Kramer,
  Eur.\ Phys.\ J.\ C {\bf 38} (2004) 93;
%
  Phys.\ Rev.\ Lett.\  {\bf 93} (2004) 232002;
%
  J.\ Phys.\ G {\bf 31} (2005) 1391;
%
  Mod.\ Phys.\ Lett.\ A {\bf 23} (2008) 1885;
%
  Eur.\ Phys.\ J.\ C {\bf 70} (2010) 91;
%
  Phys.\ Lett.\ B {\bf 508} (2001) 259;
%
  Eur.\ Phys.\ J.\ C {\bf 49} (2007) 957;
%
  Phys.\ Rev.\ D {\bf 80} (2009) 074006;

\bibitem{Guzey:2016awf}
  V.~Guzey and M.~Klasen,
  Eur.\ Phys.\ J.\ C {\bf 76} (2016) 467.

\bibitem{Guzey:2016tek}
  V.~Guzey and M.~Klasen,
  JHEP {\bf 1604} (2016) 158.

\end{thebibliography}
\end{document}